\documentclass[a4paper,11pt]{article}
\usepackage{jheppub}

\title{Asymptotic Freedom for Holographic Energy Correlators}

\author[a]{Csaba Cs\'aki,}
\author[b,c]{Ameen Ismail,}
\author[a]{and Larissa Kiriliuk}
\affiliation[a]{Laboratory for Elementary Particle Physics, Cornell University, Ithaca, NY 14853, USA}
\affiliation[b]{Enrico Fermi Institute, University of Chicago, Chicago, IL 60637, USA}
\affiliation[c]{Leinweber Institute for Theoretical Physics, University of Chicago, Chicago, IL 60637, USA}

\emailAdd{csaki@cornell.edu}
\emailAdd{ameenismail@uchicago.edu}
\emailAdd{lk559@cornell.edu}

\abstract{
We calculate energy correlators in a holographic model incorporating elements of asymptotic freedom and confinement. We model a running coupling by considering a geometry with a warp factor that deviates logarithmically from anti-de Sitter (AdS). A novel aspect of our bulk metric is that it smoothly interpolates between a Randall--Sundrum solution with a hard wall and a geometry corresponding to a logarithmic running typical of gauge theories.
By studying shockwave deformations of this metric, we compute a two-point energy correlator assuming a high-energy scalar source. This extends techniques recently developed for correlators in asymptotically AdS geometries.
We use numerical methods to find the profile of shockwaves along the extra dimension, as it does not admit an analytical form.
The running coupling leads to a decay of the two-point correlator at small angular separation, unlike the flat correlator one finds in AdS.
In the back-to-back limit we observe an exponential falloff similar to other hard-wall models.
}

\begin{document}
\maketitle
\flushbottom

\section{Introduction}
\label{sec:intro}

One of the longstanding problems in theoretical physics is to understand the dynamics of confinement in quantum chromodynamics (QCD). At high energies, QCD exhibits asymptotic freedom: the coupling constant runs to small values and perturbative computations can be performed. Below $\Lambda\sim 200$~MeV,  free quarks confine into color-neutral hadrons and the theory becomes strongly coupled. At energies well below the confinement scale one can use chiral perturbation theory, but the transition to confinement is an inherently nonperturbative phenomenon. One well-established approach is to study the dynamics of confinement through lattice simulations. A complementary technique is to study related, QCD-like theories in which one can calculate some aspects of the strong dynamics.
In particular, weakly-coupled gravitational theories in five dimensions that possess a four-dimensional, strongly-coupled dual via the anti-de Sitter / conformal field theory (AdS/CFT) correspondence~\cite{Maldacena_1999,Witten:1998qj} are a useful tool with which to study confinement~\cite{Erlich_2005,Da_Rold_2005,Sakai2005,Sakai_2005,de_T_ramond_2005,Karch_2006,Csaki:2006ji,Cs_ki_2009,ERLICH_2010}.
Such models are the focus of this work. 

Energy correlators have emerged as a simple and powerful observable to investigate confinement. An $n$-point energy correlator can be intuitively understood as measuring correlations between energy deposited in $n$ idealized calorimeters, placed far away from the interaction point in a collider experiment.
Observables of this type have been studied for many decades~\cite{PhysRevD.17.2298,Basham:1978bw,Basham:1978zq,Basham:1978}, but they have experienced a renaissance since the work of Hofman and Maldacena, who studied them in the context of conformal field theories (CFTs)~\cite{Hofman:2008ar,Hofman_2009}. Energy correlators have spurred recent developments across particle physics, ranging from collider physics~\cite{Komiske_2023,Dixon_2018,Dixon_2019,Chen_2020,Chen_2021,Chen_2022,Chen_2024,Chen222,Chen,Lee_2024,Lee2025,bossi2024,lee2025dihadronfragmentationconfinementtransition,Guo:2025zwb,chang2025quantumscalingenergycorrelators,kang2025dihadronfragmentationframeworknearside,alicecollaboration2024exposingpartonhadrontransitionjets,Firat:2023lbp,CMS:2024mlf,liu2024universalitynearsideenergyenergycorrelator,chen2025energyenergycorrelatorhadroncolliders,bhattacharya2025observableoptimizationprecisiontheory,zhao2025particlecorrelationsjets,Alipour_fard_2025,Bossi:2025nux} to formal aspects of quantum field theory~\cite{Zhiboedov_2014,Belitsky2014,Belitsky_2014,Bousso_2016,Faulkner_2016,Bousso2016,Hartman_2017,Casini_2017,C_rdova_2017,C_rdova_2018,Cordova_2018,Leichenauer_2018,Kravchuk_2018,Balakrishnan_2019,ceyhan2019recoveringqnecanec,Manenti_2020,Belin_2021,Korchemsky_2022,balakrishnan2022entropy,hartman2023averagednullenergyrenormalization,hartman2024lightraysumrulescanomaly}; for a recent review see~\cite{moult2025energycorrelatorsjourneytheory}.

In particular, the two-point energy correlator clearly illustrates the confinement transition~\cite{Komiske_2023}. In the collinear limit there is a mapping between the energy scale and the angular separation between the two idealized calorimeters~\cite{Hofman:2008ar}. At angles smaller than $\Lambda / Q$, where $Q$ is some underlying hard scale, the two-point correlator probes the low-energy theory characterized by nearly free hadrons. At larger angles it probes the high-energy theory described by asymptotically free quarks and gluons. In both of these regimes the two-point correlator can be computed analytically~\cite{Lee_2025}. However, the confinement transition between these two regimes is analytically intractable. For recent studies of the confinement transition in energy correlators, see~\cite{Lee_2024,Lee2025,bossi2024,lee2025dihadronfragmentationconfinementtransition,Guo:2025zwb,chang2025quantumscalingenergycorrelators,kang2025dihadronfragmentationframeworknearside,alicecollaboration2024exposingpartonhadrontransitionjets}.

Hofman and Maldacena also showed how to calculate energy correlators in CFTs which possess a gravitational dual through the AdS/CFT correspondence. In particular, shockwave deformations of the 5D metric encode information about energy correlators in the dual theory. These shockwaves are exact solutions to the Einstein field equations and one can superpose them to study arbitrary $n$-point energy correlators. This avoids the need to perform a laborious expansion in terms of Witten diagrams.
In light of this it is compelling to study energy correlators in holographic models of confinement~\cite{Csaki:2024joe,Csaki:2024zig}.
This was initiated in~\cite{Csaki:2024joe}, which considered an AdS extra dimension cut off by an IR brane. This is essentially a Randall--Sundrum (RS) model~\cite{Randall_1999}, dual to a strongly coupled CFT that confines at a scale set by the inverse of the brane location. The two-point correlator exhibits a transition from a constant UV regime to an exponentially falling IR regime.
This was generalized to some soft-wall models in~\cite{Csaki:2024zig}, focusing on asymptotically AdS geometries that deviate from AdS in the IR in a way that leads to confinement. Changing the 5D geometry in this way affects the form of the two-point energy correlator, but the qualitative picture remains similar.

There are a few important differences between the models studied in~\cite{Csaki:2024joe,Csaki:2024zig} and QCD. One is the UV behavior: an asymptotically AdS geometry is dual to a strongly-coupled CFT in the UV, unlike the asymptotic freedom characteristic of QCD. 
A natural next step is to study a 5D model that includes a running coupling to capture some aspects of asymptotic freedom. That is the goal of this work. Specifically, we will study the two-point energy correlator in the metric introduced in~\cite{Csaki:2006ji}, which incorporates a logarithmically running warp factor. In the UV the geometry is approximately AdS, up to logarithmic corrections. At a characteristic IR scale (the confinement scale) these logarithmic corrections blow up, leading to a singularity which cuts off the extra dimension. One new element we introduce is a parameter $\delta$, which is proportional to the anomalous dimension of the operator giving rise to the running; it modifies the dependence of the beta function on the gauge goupling. This parametrizes a family of metrics which interpolates between the RS metric with a hard cutoff and one corresponding to a gauge theory-like running coupling.

This paper is organized as follows. In Section~\ref{sec:review} we review the computation of energy correlators in CFTs with gravitational duals. In Section~\ref{sec:model} we introduce the 5D geometry that we use to model a running coupling. We demonstrate that energy correlators can be computed by studying shockwaves about the metric and construct their equation of motion in Section~\ref{sec:shockwaves}. Since an analytical solution does not exist, we then solve the shockwave equation of motion numerically in Section~\ref{sec:results} to compute energy correlators. We find the two-point correlator is substantially altered by the logarithmic running, decaying at small angular separation. The behavior in the back-to-back limit is qualitatively similar to a simple RS-like hard wall. We conclude in Section~\ref{sec:conclusions}.
Appendix~\ref{sec:wavefunction} contains technical details of the wavefunction for bulk scalar fields.

\section{Holographic energy correlators}\label{sec:review}

\subsection{Energy correlators in 4D CFTs}

We will begin with a brief review the computation of energy correlators in holography; for more details the reader is referred to~\cite{Hofman:2008ar,Csaki:2024joe}. Energy correlators are correlation functions of energy flow operators $\mathcal{E}(\vec{n})$. The energy flow operators have an intuitive interpretation as idealized calorimeters measuring the total energy deposited in a specific direction $\vec{n}$, far away from the physical process of interest. We imagine that an operator external to the CFT excites radiation comprised of CFT stuff, somewhat analogous to an $e^+ e^-$ collision producing some hadrons. This radiation propagates to future null infinity, where it is detected by the idealized calorimeters.

Our starting point is the definition of the energy flow operators in terms of the stress-energy tensor: 
\begin{equation}
   \mathcal{E}(\vec{n})= \displaystyle \lim_{r \to \infty } r^2 \int_{0}^{\infty}dt\,  T_{0i}(t,x=rn^i)n^i    \label{eq:1} .
\end{equation}
Again, this should be understood as measuring the energy deposited in a calorimeter on the celestial sphere at a location specified by the unit vector $\vec{n}$. 
This expression involves two limits, $r \rightarrow \infty$ and $t \rightarrow \infty$. An unbroken CFT is gapless and radiation propagates to future null infinity, which renders the order of these limits unimportant~\cite{Firat:2023lbp}.
We want to eventually consider gapped theories where CFT stuff propagates to future timelike infinity, so we must specify how to take these limits. Our prescription is to hold $t - r$ constant while sending $t + r \rightarrow \infty$~\cite{Belitsky2014,Chen_2020}.

It is more convenient to perform computations in lightcone coordinates 
\begin{equation}
    x^{\pm}=(t-r) + r(1\pm \cos \theta), \quad x^{1}+ix^2=re^{i\phi} \label{eq:2} .
\end{equation}
Then the energy flow operators take the form
\begin{equation}
    \mathcal{E}(\vec{n})= \displaystyle \lim_{x^+ \to \infty } \frac{(x^{+})^{2}}{4} \int_{-\infty}^{\infty}dx^-\,  T_{--}(x^+,x^-,x^\perp),\label{eq:3}
\end{equation}
with $x^\perp =x^{1,2}$. We can avoid taking the limit $x^+ \to \infty$ by making a conformal transformation  
\begin{equation}
   x^+\to -\frac{\ell^2}{x^+} , \quad x^-\to x^- - \frac{\lvert x^\perp \rvert^2}{x^+}, \quad x^\perp \to \ell \frac{x^\perp}{x^+} \label{2.4}
\end{equation}
which maps the celestial sphere at $x^+ = \infty$ to the null plane $x^+ = 0$.
Specifically, a point $(\theta, \phi)$ on the sphere is mapped to $x^\perp$ on the null plane such that $x^1 + i x^2 = e^{i\phi} \tan \theta/2$.

The energy correlator then takes a simple form on the plane:
\begin{equation}
    \mathcal{E}(\vec{n})= \left(1+ \lvert x^\perp \rvert^2  \right)^3\int_{-\infty}^{\infty}dx^-\,  T_{--}(x^+=0,x^-,x^\perp),\label{eq:2.5}
\end{equation}
where the factors of $1+ \lvert x^{\perp} \rvert^2$ can be derived from the Jacobian of the coordinate transformation.

We will mainly perform computations on the $x^+=0$ plane for the sake of simplicity. These can always be mapped back to the celestial sphere. 

\subsection{Energy correlators in AdS}

Let us now consider a CFT with a holographic dual. We will first work in AdS, which is dual to an unbroken CFT, following~\cite{Hofman:2008ar}. We will then review the generalization to 5D models exhibiting confinement~\cite{Csaki:2024joe,Csaki:2024zig}.

The external operator that excites the CFT is dual to a bulk field sourced on the boundary of AdS. Each insertion of an energy flow operator sources a bulk graviton.
Hence an $n$-point energy correlator corresponds to an $n+2$-point correlation function involving two bulk fields (for the source) and $n$ bulk gravitons.
 
In principle we could calculate energy correlators by directly evaluating the relevant Witten diagrams, but this would be fairly laborious.
We can compute them more easily by studying shockwave deformations of the metric, which are the gravitational duals of exponentiated energy flow operators~\cite{Hofman:2008ar}. Linear superpositions of shockwaves constitute exact solutions to the Einstein equations, which allows us to easily insert $n$ energy flow operators in the path integral to compute an $n$-point energy correlator. Ultimately, we will compute energy correlators by studying the two-point function of a bulk field propagating in a shockwave geometry.

Recall the AdS metric in Poincar\'e coordinates is
\begin{equation}
    ds_{\rm AdS}^2 = \frac{R^2}{z^2} \left( dx^+ dx^- - \lvert dx^\perp \rvert^2 - dz^2 \right) .
\end{equation}
The AdS boundary lies at $z = 0$. Analogously to Eq.~\eqref{2.4}, one can perform a coordinate transformation mapping the celestial sphere to the null plane $x^+ = 0$:
\begin{equation}
   z \rightarrow R \frac{z}{x^+}, \quad x^- \rightarrow x^- - \frac{\lvert x^\perp \rvert^2 + z^2}{x^+}, \label{eq:transformation}
\end{equation}
and $x^+, x^\perp$ transform as in Eq.~\eqref{2.4} with $\ell = R$.
In what follows we set $R = 1$.

The insertion of an exponentiated energy flow operator at $x^\perp = y^\perp$, $\exp \epsilon \mathcal{E}(y^\perp)$, sources a metric perturbation referred to as a shockwave:
\begin{equation}
    ds^2 = ds^2_{\textrm{AdS}} + \frac{\epsilon}{z^2} \delta(x^+)f(x^{\perp} - y^\perp, z) \; (dx^+)^2 .
\end{equation}
The function $f(x^\perp,z)$ describes the profile of the shockwave in the extra dimension. The shockwave is localized at $x^+ = 0$; it satisfies a boundary condition $f(x^\perp, z=0) = \delta^2(x^\perp)$, corresponding to where we insert the energy flow operator. The Einstein equations for the metric with a shockwave reduce to a linear differential equation for the profile,
\begin{equation}
    \left [ \frac{3}{z}\partial_{z} -\left(\partial_{1}^2+\partial_{2}^2+\partial_{z}^2 \right) \right] f(x^\perp, z) =0 .
\end{equation}
The solution consistent with the boundary condition and bounded as $z \rightarrow \infty$ is $f(x^\perp, z) = z^4 / (z^2 + \lvert x^\perp \rvert^2)^3$.
Because this equation is linear, we can describe multiple insertions of energy flow operators by superposing shockwaves at different points on the null plane. This makes it easy to compute energy correlators. 

The final ingredient is to specify the external source that excites the CFT.
For simplicity we will assume a scalar source with four-momentum $q^{\mu}=(q,\vec{0})$, which is the appropriate physical setup for a collision in the center-of-momentum (COM) frame.
This sources a bulk scalar field $\phi$ with a plane wave boundary condition. By studying the wavefunction of the scalar in the presence of shockwaves, we can extract energy correlators. The presence of the shockwave leads to a discontinuity in $\phi$ at the null plane $x^+ = 0$, which can be seen from the equation of motion
\begin{equation}
    \partial_{-}\partial_{+} \phi + \epsilon \delta(x^+) f(x^{\perp},z) \partial_{-}^2 \phi =0
\end{equation}
There are other terms in the equation of motion for $\phi$, but they are negligible at $x^+ = 0$.
Integrating over the discontinuity yields a jump condition for the scalar field
\begin{equation}
    \partial_- \phi(x^+ \rightarrow 0^+)=e^{-\epsilon f\,\partial_{-}} \partial_-\phi(x^+ \rightarrow 0^-) ,
\end{equation}
essentially causing a shift in $x^-$. This is for a single shockwave; for multiple shockwaves a similar equation holds with a shift for each shockwave.
Away from the discontinuity the scalar field profile is just given by the AdS wavefunction. In AdS the wavefunction evaluated at the boundary $x^+=0$ takes the form $\phi\sim e^{iqx^-}\delta^{2}(x^{\perp})\delta(z-1)$\footnote{This is easiest to show by embedding 5D AdS in 6D Euclidean space~\cite{Hofman:2008ar}.} .

We can then compute the holographic action and expand at leading order in the shockwaves to extract correlation functions, e.g.
\begin{equation}
    \langle e^{\epsilon \mathcal{E}(y^\perp)} \rangle \sim \int \frac{dz}{z^3} d^2 x^\perp d x^- i \phi^* \exp \left[ -\epsilon \left(1 + (y^\perp)^2 \right)^3 f(x^\perp - y^\perp, z) \partial_- \right] \partial_- \phi \Big |_{x^+ = 0} + \rm{~ c.c.}
\end{equation}
for a one-point function.
The integration is trivial because the wavefunction is delta-function localized. For a scalar source we can place one of the energy flow operators at $y^\perp = 0$ without loss of generality. Thus, the two-point energy correlator can be written as
\begin{equation}
     \langle \mathcal{E}(0)\mathcal{E}(y^{\perp}) \rangle \sim \left ( 1+(y^{\perp})^2 \right )^3 f(y^{\perp},z=1) . \label{2.10}
\end{equation}
There is an overall normalization related to the total energy of the source which we are not keeping track of.

Since the shockwave is $f(x^\perp, 1) = (1+(y^{\perp})^2)^{-3}$, the correlator is a constant, $ \left< \mathcal{E}\mathcal{E}\right> \sim 1$. This is exactly what one expects, as the dual theory is a strongly-coupled CFT. There is no scale or preferred direction, leading to an isotropic, featureless distribution --- a ``mush'' of energy spread evenly across all angles. This will change when we modify the geometry to model confinement. The introduction of a scale leads to an angular dependence in the energy correlator.

\subsection{Introducing confinement}
The computation of holographic energy correlators was extended to 5D models exhibiting confinement in~\cite{Csaki:2024joe,Csaki:2024zig}. The simplest possibility is to cut off AdS with an IR brane at $z = z_{\rm IR}$~\cite{Csaki:2024joe}, similar to a Randall--Sundrum (RS) model~\cite{Randall_1999}. In~\cite{Csaki:2024zig} this was generalized to soft-wall models of confinement with a metric
\begin{equation}
    ds^2 = e^{-2A(z)} \left( dx^+ dx^- - \lvert dx^\perp \rvert^2 - dz^2 \right) .
\end{equation}
The warp factor $A(z) \sim \log z/R$ as $z \rightarrow 0$ for the metrics in~\cite{Csaki:2024zig}, i.e. the geometry is asymptotically AdS.

Importantly, the coordinate transformation in Eq.~\eqref{eq:transformation} needs to be modified in order to remain an isometry of the metric. Since the CFT is spontaneously broken, there is an explicit length scale in the metric corresponding to the (inverse) scale of confinement.
If we treat this scale as a spurion that also transforms appropriately, then Eq.~\eqref{eq:transformation} remains an isometry.
For instance, in the RS case the scale is the brane location $z_{\rm IR}$, and it must transform as $z_{\rm IR} \rightarrow z_{\rm IR} / x^+$.

In the RS case, one trivially has linear shockwave solutions to the Einstein equations, since the metric in the bulk is just AdS. 
As shown by~\cite{Csaki:2024zig}, an arbitrary warped geometry also admits linear shockwave solutions.
Therefore, it is possible to superpose them to compute $n$-point energy correlators. If the extra dimension is infinite, as in AdS, then we require the shockwave does not diverge at $z \rightarrow \infty$. If the extra dimension is cut off by a brane or singularity at $z_{\rm IR}$ (like the RS case), we impose a Neumann boundary condition:
\begin{equation}
    \partial_{z}f(x^{\perp},z) \Big |_{z=z_{IR}}=0 .
\end{equation}

Since the 5D geometry is modified, the wavefunctions for bulk fields will be different. In general this would make it difficult to compute energy correlators, as the computation is greatly simplified by the scalar wavefunction being localized in the extra dimension, $\phi \sim \delta(z-1)$. However, for a bulk scalar with energy well above the scale of confinement ($q \gg 1/z_{IR}$), the wavefunction is insensitive to the low-energy dynamics of confinement. So in this limit we recover the AdS wavefunction and we can compute energy correlators.

Recall that in the definition of the energy flow operator there are two limits, one associated with the integral over time and the other from sending the idealized calorimeter to future null infinity. In a gapped theory one must be careful about the order of these limits, since massive particles flow to future timelike infinity rather than future null infinity. Ultimately, one finds the two-point energy correlator is still given by Eq.~\eqref{2.10} (for a high-energy scalar source)~\cite{Csaki:2024joe}.
Modifying the geometry changes the profile of shockwaves and therefore changes energy correlators.
For example, in the RS case the IR brane causes the two-point energy correlator to exhibit an exponential decay for $\lvert x^\perp \vert > z_{\rm IR}$.
This is not the same scaling as observed in QCD energy correlators~\cite{moult2025energycorrelatorsjourneytheory}, but it still indicates confinement, albeit via a different mechanism than in QCD. The RS case corresponds to a hard-wall type of confinement, since AdS is abruptly cut off at $z=z_{IR}$.

\section{5D model with running}\label{sec:model}

We would like to consider a metric which deviates from AdS to incorporate some aspects of asymptotic freedom. While we will mostly follow the discussion in~\cite{Csaki:2006ji}, we will introduce one additional parameter that will allow us interpolate between the RS limit and a running coupling. 

We introduce a scalar field $\Phi$ which propagates in the 5D bulk. The action is
\begin{equation}
    \int d^5 x \sqrt{g} \left[ -\frac{1}{2\kappa^2} \mathcal{R} + \frac{3}{2\kappa^2} g^{ab} \partial_a \Phi \partial_b \Phi - V(\Phi)  \right].
\end{equation}
The unusual normalization of the $\Phi$ kinetic term is chosen for future convenience; the canonical field is $\phi = \sqrt{3} \Phi/\kappa$. The goal is to properly choose the scalar potential so as to model a running coupling. We parametrize the metric as
\begin{equation}
    ds^2 = e^{-2A(y)}\eta_{\mu\nu} dx^\mu dx^\nu - dy^2 = e^{-2A(v)} \eta_{\mu\nu} dx^\mu dx^\nu - \frac{R^2}{v^2}dv^2 ,
\end{equation}
where $A$ is the warp factor. These two choices for the coordinate along the fifth direction are related by $v/R = e^{y/R}$. We emphasize that $v$ is \textit{not} the same as the conformal coordinate $z$, defined by $dz/dy = e^{A}$. The two coincide only when the metric is exactly AdS.

We use the superpotential method to find a solution to the coupled Einstein-scalar equations~\cite{DeWolfe:1999cp}. Given any solution that depends only on the fifth coordinate $y$, one can define a superpotential $W[\Phi]$ via the equations
\begin{equation}\label{eq:superpotential}
    \frac{d\Phi}{dy} = W'[\Phi], \quad \frac{dA}{dy} = W[\Phi] .
\end{equation}
The superpotential must satisfy the consistency condition $6 V(\Phi)/\kappa^2 =W'^2/4 -W^2$.
As in~\cite{Csaki:2006ji}, we assume that the gauge coupling $g$ of the CFT is related to the exponential of the scalar field $\Phi$ in the 5D theory: $\alpha = g^2/4\pi \leftrightarrow e^{2\Phi}$. We identify the coordinate $v$ with the inverse of the running scale $\mu^{-1}$. Since the running of the gauge coupling in QCD (and general asymptotically free gauge theories in 4D) is given by $\alpha^{-1}(\mu )=\beta_0 \log \frac{\mu}{\Lambda}$, we take the following ansatz for the scalar profile:
\begin{equation}\label{eq:ansatz}
    e^{2\Phi} = \frac{1}{ \beta_0 \log v_0/v} = \frac{ R}{\beta_0 (y_0 - y)} .
\end{equation}
The blowing up of the QCD coupling at $\Lambda$ is modeled via a singularity at
$v_0 , y_0$ along the extra dimension, implying $v_0 = \Lambda^{-1}$. Since $\Lambda =\mu  e^{-1/\beta_0 \alpha(\mu)}$, we expect that the effect of increasing $\beta_0$ is to reduce the distance to the singularity (for a fixed value of the coupling in the UV). That is, $v_0(\beta_0) \propto \tilde{v}_0 \lambda^{1/\beta_0}$, where $\tilde{v}_0$ and $\lambda$ are constants. The scalar solution and the metric for this ansatz are then obtained by solving Eq.~\eqref{eq:superpotential}:
\begin{equation}
    A(y)=\frac{y}{R} -\frac{1}{4} \log \left(1-\frac{y}{y_0} \right)\ ,
\end{equation}
where we choose the constant such that $A(0)=0$. The superpotential is given by
\begin{equation}
    W[\Phi ]= \frac{\beta_0}{4R} e^{2\Phi} +W_0 ,
\end{equation}
where the constant is $W_0 = 1/R$.
Note that while the metric itself does not directly depend on $\beta_0$, the geometry implicitly does via the $\beta_0$-dependence of $y_0$. In the $\beta_0 \rightarrow 0$ limit, the singularity goes to infinity, $y_0 \rightarrow \infty$, and the metric becomes AdS. This is expected since when the beta function vanishes we have a truly conformal theory.

To smoothly interpolate from this solution to the RS hard-wall solution, we introduce another parameter $\delta$ into our ansatz, replacing Eq.~\eqref{eq:ansatz} with the modified version
\begin{equation}\label{eq:ansatzwithdelta}
    e^{2\Phi/\delta} = \frac{\delta}{ \beta_0 \log v_0/v} = \frac{\delta R}{\beta_0 (y_0 - y)} .
\end{equation}
One can gain some intuition about the physical meaning of $\delta$ by calculating the scaling dimension  $\Delta$ of the operator associated with the bulk scalar: 
\begin{equation} \label{scaling}
    \Delta(v)=\frac{d\log e^{\Phi}}{d\log v}=\frac{d\log e^{\log (\delta/2 \log v/v_0 )}}{d\log v}=\frac{\delta}{2 \log v_0/v} .
\end{equation}
This shows that $\delta$ controls the scaling dimension of the corresponding operator. It is also instructive to calculate the modification of the holographic beta function due to $\delta$:
\begin{equation}
    -\frac{d\alpha}{d\log v}=-\frac{d}{d\log v}\left [ \frac{\delta}{\beta_0 \log v_0/v} \right ]^{\delta} = \beta_0 \left [\frac{\delta}{\beta_0 \log v_0/v} \right ]^{1+\delta}=\beta_0 \alpha^{1+1/\delta} ,
\end{equation}
which shows that $\delta$ modifies the scaling of the beta function with $\alpha$. For $\delta =1$ we reproduce the usual running expected in QCD. Interestingly, in the limit $\delta\to 0$ the scaling dimension $\Delta$ vanishes everywhere except at the singularity $v=v_0$. This corresponds to a hard wall RS-type scenario, wherein the theory is exactly conformal until an anomalous dimension suddenly blows up, triggering confinement and chiral symmetry breaking. 

Let us now derive the metric.  We use Eqs.~\eqref{eq:ansatzwithdelta} and \eqref{eq:superpotential} to calculate the superpotential, leading to
\begin{equation}
    W[\Phi] = \frac{\beta \delta}{4 R} e^{2 \Phi/\delta} + W_0 ,
\end{equation}
where $W_0$ is an integration constant.
We set $W_0 = 1/R$ to obtain an AdS-like metric with curvature scale $R$. Imposing $A(0)=0$, the warp factor is given by
\begin{equation}\label{eq:warpfactor}
    A = \frac{y}{R} - \frac{\delta^2}{4} \log \left(1 - \frac{y}{y_0} \right) = \log \frac{v}{R} - \frac{\delta^2}{4} \log \left[ 1 - \frac{\log v / R}{\log v_0/R} \right] .
\end{equation}
Note that this is the same metric as derived in Ref.~\cite{Csaki:2006ji}, up to a constant shift of the warp factor.
Explicitly, the line element is
\begin{equation}\label{eq:metric}
    ds^2 = \left( \frac{R}{v} \right)^2 \left[ \left(1 - \frac{\log v/R}{\log v_0/R} \right)^{\delta^2/2} dx^\mu dx_\mu - dv^2 \right] .
\end{equation}
Taking the limit $\delta \rightarrow 0$ we obtain AdS cut off by a brane at $v = v_0$, which is just RS, in agreement with our expectations.

In the remainder of this paper we study energy correlators in the model described by Eq.~\eqref{eq:metric}. Before moving on, it is useful to write down an explicit expression for the conformal coordinate $z$, defined by $dz/dy = e^{A(y)}$:
\begin{equation}\label{eq:conformalcoordinate}
    z = R e^{y_0 / R} \left( \frac{y_0}{R} \right)^{\delta^2/4} \Gamma \left(1 - \frac{\delta^2}{4}, \frac{y_0-y}{R} \right) = v_0 \left( \log \frac{v_0}{R} \right)^{\delta^2/4} \Gamma \left(1 - \frac{\delta^2}{4}, \log \frac{v_0}{v} \right)
\end{equation}
where $\Gamma$ is the upper incomplete gamma function\footnote{The upper incomplete gamma function is defined by $\Gamma(a, x) = \int_x^\infty dt \: t^{a-1}e^{-t}$.}.
Note that we choose the integration constant such that $y \rightarrow -\infty$ corresponds to $z = 0$. Also, the singularity lies at $z_0 = v_0  (\log v_0/R)^{\delta^2/4} \Gamma(1 - \delta^2/4)$.

\section{Shockwaves in the model}\label{sec:shockwaves}

We are interested in computing holographic energy correlators with the metric in Eq.~\eqref{eq:metric}. In what follows we set $R = 1$. We shall compute the two-point correlator for a scalar source with energy well above the confinement scale.

In~\cite{Csaki:2024zig}, the formalism for computing energy correlators was generalized to asymptotically AdS geometries. Most of those results carry over to our case, even though our warp factor is not asymptotically AdS. Specifically, three things need to be shown: 
\begin{itemize}
    \item One can perform the coordinate transformation in Eq.~\eqref{eq:transformation} mapping the celestial sphere to the null plane.
    
    \item Shockwaves about the metric obey a linear equation of motion.
    
    \item The wavefunction for a scalar source is delta-function localized in the high-energy limit.
\end{itemize}
As noted in~\cite{Csaki:2024joe}, the first of these conditions is satisfied as long as one treats the IR scale $z_0$ as a spurion transforming as $z_0 \rightarrow z_0 / x^+$. Additionally,~\cite{Csaki:2024zig} demonstrated that shockwaves are linear in an arbitrary warped metric, and in particular that there is no mixing between between the shockwave and fluctuations of the bulk scalar which stabilizes the geometry. All that remains is to study the wavefunction for a high-energy scalar source, which we do in Appendix~\ref{sec:wavefunction}. Just as in the soft-wall scenarios discussed in~\cite{Csaki:2024zig}, we find the wavefunction is localized in the extra dimension as $\delta(z - 1)$, where $z$ is the conformal coordinate. (Remember that the conformal coordinate $z$ is not the same as $v$.)

The upshot is that the two-point energy correlator is simply given in terms of the shockwave by Eq.~\eqref{2.10}.

This model has a hard wall at $v = v_0$, suggesting a similar mechanism of confinement to the RS case. Intuitively, we expect that at $x^\perp > v_0$ the two-point energy correlator is similar to the RS two-point correlator. The metric is not AdS-like in the UV, so we expect to see some novel features at small $x^\perp$. These aspects will be made explicit by our numerical results in Section~\ref{sec:results}.

\subsection{Equation of motion} 
We parametrize the shockwave about our metric as
\begin{equation}
    ds^2= e^{-2A(v)} \left( dx^+dx^- -\left (dx^\perp \right)^2 \right)-\frac{dv^2}{v^2}+ \epsilon e^{-2A(v)} \delta(x^+)f(x^{\perp},v) \left(dx^+ \right)^2 .
\end{equation}
Substituting this metric into the Einstein equations gives the equation of motion for the shockwave $f(x^\perp, v)$. In doing so one must be careful to include the effect of the shockwave on both the Einstein tensor and the stress-energy tensor for the bulk scalar field. The latter is affected by the shockwave because $T_{\mu\nu} \supset -\mathcal{L} g_{\mu\nu}$.

We find the equation of motion
\begin{equation}\label{eq:shockwaveEOM}
    \left[ \left( \frac{\delta^2}{\log v_0/v} + 3 \right) \frac{1}{v} \partial_v - \partial_v^2 - \left(1 - \frac{\log v}{\log v_0} \right)^{-\delta^2/2} \partial_\perp^2 \right] f(x^\perp, v) = 0 .
\end{equation}
The shockwave is subject to the usual UV boundary condition $f(x^\perp, v = 0) \sim \delta^2(x^\perp)$, corresponding to a localized insertion of $T_{--}$ on the null plane $x^+ = 0$. Since the extra dimension is cut off at a finite proper distance as in RS, we should impose a Neumann boundary condition at the singularity: $\partial_v f(x^\perp, v = v_0) = 0$.

The shockwave equation of motion can be expressed in a Schr\"odinger-like form by rescaling the shockwave and Fourier transforming over the perpendicular directions $x^\perp$:
\begin{equation}
    g(v, k^\perp) = v^{-3/2} \left(1 - \frac{\log v}{\log v_0} \right)^{\delta^2/2} \int d^2 x^\perp e^{i k^\perp \cdot x^\perp} f(x^\perp, v) .
\end{equation}
This gives an equation $(\partial_v^2 - U(v) ) g = 0$, with the potential
\begin{equation}\label{eq:schrodingerEOM}
    U(v) = \frac{1}{v^2} \left[ \frac{15}{4} + \frac{2 \delta^2}{\log v_0/v} - \frac{\delta^2}{2} \left(1 - \frac{\delta^2}{2} \right) \frac{1}{\log^2 v_0/v}  \right] + \left(1 - \frac{\log v}{\log v_0} \right)^{-\delta^2/2} k^2 .
\end{equation}
In the limit $\delta \rightarrow 0$ or $v_0 \rightarrow \infty$, the metric is the same as AdS and the potential reduces to $15/4v^2 + k^2$, in agreement with~\cite{Csaki:2024joe}.

The equation of motion in Eq.~\eqref{eq:shockwaveEOM} does not admit an analytical solution, so in Section~\ref{sec:results} we will solve it numerically. Terms of the form $1/ \log v_0/v$ blow up near the singularity, so the shockwaves exhibit a boundary layer behavior. Our strategy will be to construct an analytical solution in the boundary layer and stitch it to a numerical solution away from the singularity.
Away from the singularity, one can also consider a WKB approximation for the shockwave. However, the approximation breaks down near $v = 0$ due to the presence of a turning point, so it is not very useful for computing energy correlators.

\subsection{Boundary layer solution}
Let us solve the equation of motion near the singularity. We take Eq.~\eqref{eq:shockwaveEOM} and change variables to $w = 1 - v/v_0$. Expanding about the singularity at $w = 0$, the leading-order equation of motion is
\begin{equation}\label{eq:boundaryeom}
    \left[ \partial_w^2 + \frac{\delta^2}{w} \partial_w - \left( \frac{\log v_0}{w} \right)^{\delta^2/2} \ell^2 \right] f(w, \ell) = 0
\end{equation}
where $\ell = k v_0$.
This can be solved in terms of Bessel functions, resulting in  
\begin{equation}
    f(w,\ell)= w^{(1-\delta^2)/2} \left [ c_1(\ell)K_\nu\left ( \frac{(\log v_0)^{1-p}}{p} \ell w^{p} \right ) +c_2(\ell) I_{\nu}\left (   \frac{(\log v_0)^{1-p}}{p} \ell w^{p}\right )      \right ] ,
\end{equation}
with 
\begin{equation}
    p = 1 - \frac{\delta^2}{4}, \quad \nu = \frac{\lvert \delta^2-1 \rvert}{2-\delta^2/2} .
\end{equation}
Recall that $f$ satisfies a Neumann BC at the singularity, so $\partial_w f = 0$ at $w = 0$. This immediately gives $c_1(\ell) = 0$, leading to
\begin{equation}
    f(w,\ell) = c(\ell) w^{(1-\delta^2)/2}  I_{\nu}\left (   \frac{(\log v_0)^{1-p}}{p} \ell w^{p}\right ). \label{eq:boundarysolution}
\end{equation}
The coefficient $c(k)$ will be fixed by matching to a numerical solution away from the singularity.

We remark that there is a subtlety in taking the RS limit $\delta \rightarrow 0$ in the above discussion. For arbitrarily small but nonzero $\delta$ there is a boundary layer at $v_0(1 - \delta^2) \lesssim v < v_0$, where the shockwave is described by Eq.~\eqref{eq:boundarysolution}. When $\delta = 0$ exactly, there is no boundary layer. Correspondingly, Eq.~\eqref{eq:boundaryeom} ceases to be the correct expansion of the equation of motion near the singularity.

\section{Numerical results}\label{sec:results}

We numerically solve the equation of motion for the shockwave (see Eq.~\eqref{eq:shockwaveEOM}), working in Fourier space in the transverse coordinates $x^\perp$. We impose a Dirichlet BC on the UV brane, $f(v=0, k^\perp) \sim 1$, which is the Fourier transform of the delta-function boundary condition $f(v = 0, x^\perp) \sim \delta^2(x^\perp)$. We then match $f$ and its first derivative $\partial f / \partial v$ to the analytical solution near the singularity derived in Eq.~\eqref{eq:boundarysolution}. The matching procedure is expected to introduce an error of order $1 - v_{m}/v_0$, where $v_m$ is the point at which we perform the matching. We have verified that our results are insensitive to variations in the matching point $v_m$, provided it is sufficiently close to the singularity.

\begin{figure}[htbp]
    \centering
    \includegraphics[width=0.8\textwidth]{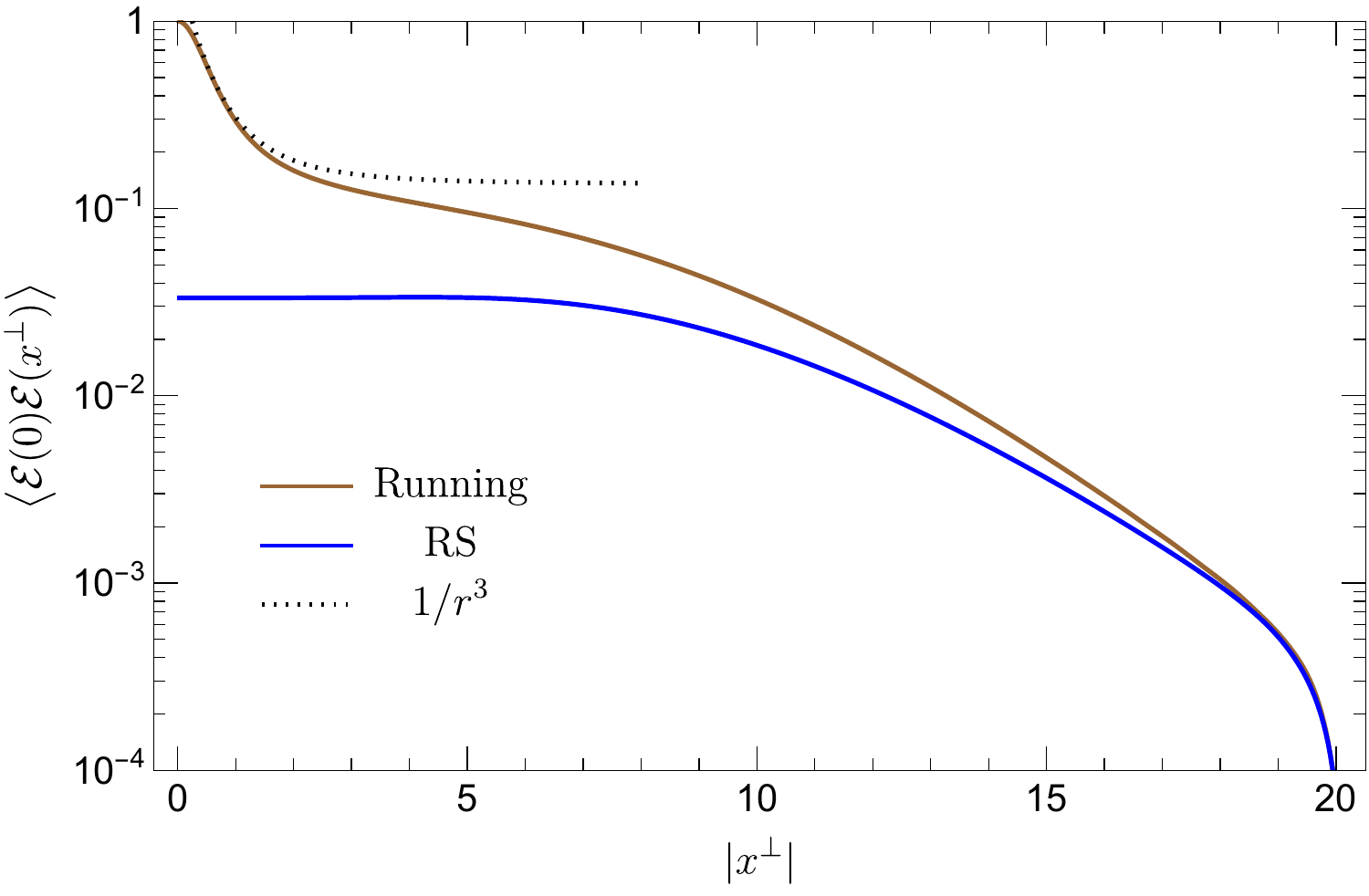}
    
    \caption{%
        The two-point energy correlator as a function of distance on the transverse plane $r = \lvert x^{\perp} \rvert$ (note the log scale). 
        The burgundy curve corresponds to the model with a running coupling. For comparison we show the correlator for a simple hard wall (RS) model in blue. The normalization is arbitrary. We choose to set the correlators to be equal at a point at large $\lvert x^\perp \rvert$ for visualization purposes. The black dashed line shows a fit of the correlator for the running model to a $1/r^3$ power law.}
    \label{fig:sameconf}
\end{figure}

In Fig.~\ref{fig:sameconf} we present our main result: the two-point energy correlator in our model as a function of separation on the null plane $\lvert x^\perp \rvert$. We assume a high-energy scalar source and fix $\delta = 1$, $v_0 = 5$. For comparison, we also show the two-point energy correlator in a simple hard-wall (RS) model with the same brane location $v_{0}=5$. Note that the overall normalization is arbitrary. We normalize the correlators to be equal at a point at large $\lvert x^\perp \rvert$.

As seen in Fig.~\ref{fig:sameconf}, the energy correlator in the running model is strikingly different from RS at small $x^\perp$, corresponding to the collinear limit on the celestial sphere. We find that the correlator approximately follows a power law around $\lvert x^\perp \rvert = 1$. We have not obtained an analytical explanation of this scaling, but this would be an interesting direction for future inquiry. In Fig.~\ref{fig:sameconf} we fit the correlator between $\lvert x^\perp \rvert  = 0.5$ and $\lvert x^\perp \rvert = 2$ to a $1/r^3$ power law. Very close to $x^\perp = 0$, the correlator flattens out and is not described by this power law. 

Both correlators behave similarly at large $\lvert x^{\perp} \rvert$ (corresponding to the back-to-back limit), falling off exponentially. This is expected given that in both cases the extra dimension is cut off at a finite proper distance. It suggests that the underlying confinement mechanism is similar in the two models. This is further corroborated by the fact that RS can be obtained from the running model in the $\delta \rightarrow 0$ limit; we will study energy correlators in this limit below.

\begin{figure}
  \hspace*{-1cm} 
  \includegraphics[scale=0.42]{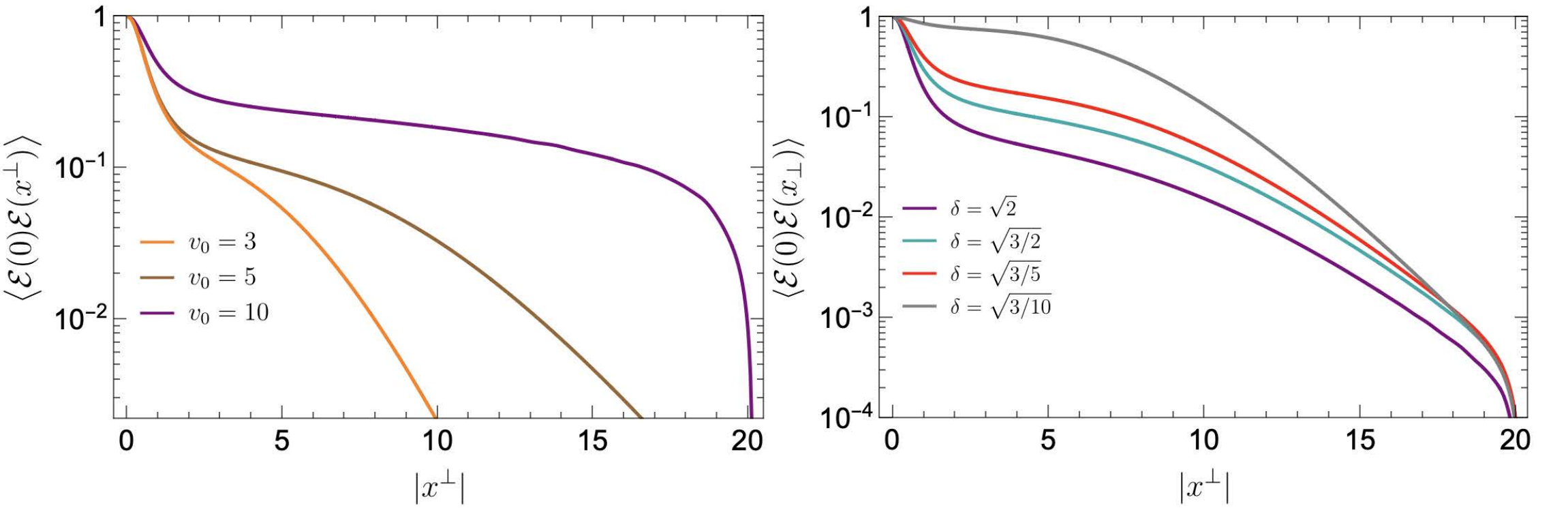}
  
  \caption{The effect of varying the model parameters on the two-point energy correlator. In the left panel we show the correlator for three different locations of the IR brane $v_0 =3, 5, 10$, holding $\delta = 1$ fixed. In the right panel we show the correlator for $\delta^2 = 2, 3/2, 3/5, 3/10$, with a fixed brane location $v_0=5$. We choose the normalization such that the correlator is one at $x^\perp = 0$.}
  
  \label{fig:main}
\end{figure}

Fig.~\ref{fig:main} illustrates the effect of varying the model parameters $v_0$ and $\delta$ on the two-point energy correlator. In the left panel we hold $\delta = 1$ fixed and show the energy correlator for different locations of the singularity, $v_0 = 3, 5, 10$. As $v_0$ is increased, the IR regime (where the correlator falls exponentially) begins at larger values of $\lvert x^\perp \rvert$. In the limit $v_0 \rightarrow \infty$ the metric is just AdS and we expect to recover a constant correlator.
In the right panel of Fig.~\ref{fig:main} we hold $v_0 = 5$ fixed and show the effect of varying $\delta$. Recall that $\delta$ is related to the scaling dimension of a CFT operator  whose VEV is responsible for breaking conformal symmetry, and that the limit $\delta \rightarrow 0$ corresponds to RS. Indeed, as we reduce $\delta$, the energy correlator gradually flattens out at small $\lvert x^\perp \rvert$, approaching the RS energy correlator~\cite{Csaki:2024joe}.

It is noteworthy that our results deviate from a constant, scale-invariant correlator in both the regimes of small and large $\lvert x^\perp \rvert$. This is in contrast to the RS and soft-wall models studied in~\cite{Csaki:2024joe,Csaki:2024zig}, which only exhibit such behavior for large $\lvert x^\perp \rvert$. The difference is that those were all asymptotically AdS geometries, dual to an IR deformation of the CFT, while our geometry also differs from AdS in the UV. We believe this is responsible for the behavior of the two-point correlator at small $\lvert x^\perp \rvert$.

Lastly, recall that we can map the separation on the transverse plane back to an angular separation on the celestial sphere, using $\lvert x^\perp \rvert = R \tan \theta/2$. From this perspective, what we are observing is a modification of the two-point correlator in the collinear regime.

\section{Conclusions}\label{sec:conclusions}
In this paper we have studied energy correlators in a holographic model incorporating a running coupling. This was possible because linear shockwave solutions exist about a geometry with an arbitrary warp factor. We modeled running by including a logarithmic term in the warp factor, which also leads to a singularity indicating confinement. 

We calculated the two-point energy correlator in this model, assuming a high-energy scalar source. This generalizes the procedure developed for computing holographic energy correlators in confining theories in~\cite{Csaki:2024joe,Csaki:2024zig}.
Our main result is that the two-point correlator falls as a power law for small separation $\lvert x^\perp \rvert$. Interestingly, a similar feature is seen in energy correlators in QCD-like theories \cite{moult2025energycorrelatorsjourneytheory}. This behavior is unlike AdS with a hard-wall cutoff or asymptotically AdS soft-wall models. We attribute it to the inclusion of a logarithmic running in our warp factor. In the IR regime, at $\lvert x^\perp \rvert \gg v_0$, we observed an exponential falloff, similar to what one sees in other confining holographic models. This appears to be a generic feature of holographic confinement, as explored in ~\cite{Csaki:2024zig,Csaki:2024joe}. 

A novel feature of our holographic model is a parameter $\delta$, which defines a family of metrics smoothly interpolating between a gauge theory-like running coupling and an RS-like hard wall. We related this parameter to the scaling dimension $\Delta$ of the bulk scalar field that triggers confinement. The $\delta \rightarrow 0$ limit reproduced RS, corresponding to a theory which is exactly conformal above the IR scale. We computed the two-point energy correlator for different values of $\delta$. In accordance with our interpretation of $\delta$, the result approached the RS energy correlator in the $\delta \rightarrow 0$ limit.

Our work represents an important step in understanding QCD energy correlators using holography. These results show that features in the collinear limit of the energy correlator arise when one implements a running coupling in the holographic model.  

A natural next step would be to incorporate jets by computing the leading-order stringy corrections. This was done for the AdS case by~\cite{Hofman:2008ar}, but has not been studied in holographic models with confinement. Another interesting direction would be to consider sources other than scalars.

\acknowledgments

CC and LK are supported in part by the NSF grant PHY-2309456. CC is also supported in part by the US-Israeli BSF grant 2024091. AI is supported by a Mafalda and Reinhard Oehme Postdoctoral Research Fellowship from the Enrico Fermi Institute at the University of Chicago. CC and AI also thank the Aspen Center for Physics (supported by National Science Foundation grant PHY-2210452) and the Munich Institute for Astro, Particle and Biological Physics (funded by the DFG grant EXC-2094 --- 390783311) for their hospitality while this work was in progress.  

\appendix

\section{Scalar wavefunction\label{sec:wavefunction}}

Here we study the wavefunction for a bulk scalar field in our metric, which we reproduce below in terms of the coordinates $v$ and $z$ (the conformal coordinate):
\begin{equation}
    ds^2 =  e^{-2A(v)} \eta_{\mu\nu} dx^\mu dx^\nu - \frac{1}{v^2}dv^2 = e^{-2A(z)} \left( \eta_{\mu\nu} dx^\mu dx^\nu - dz^2 \right).
\end{equation}
The warp factor $A(v)$ was given in Eq.~\eqref{eq:warpfactor}, and we work in units where $R = 1$.

The conformal coordinate is given in terms of $v$ by Eq.~\eqref{eq:conformalcoordinate}, reproduced here:
\begin{equation}
    z(v) = v_0 \left( \log v_0 \right)^{\delta^2/4} \Gamma \left(1 - \frac{\delta^2}{4}, \log \frac{v_0}{v} \right) .
\end{equation}
It will be useful to evaluate $z'(v)$ and $z''(v)$; we find
\begin{equation}\label{eq:zderivatives}
    z'(v) = \left(1 - \frac{\log v}{\log v_0} \right)^{-\delta^2/4}, \quad z''(v) = \frac{\delta^2}{4 v \log v_0/v} \left(1 - \frac{\log v}{\log v_0} \right)^{-\delta^2/4} .
\end{equation}

We consider the equation of motion for a bulk scalar field $\phi$:
\begin{equation}
    \left[ \left( \frac{\delta^2}{\log v_0/v} + 3 \right) \frac{1}{v} \partial_v - \partial_v^2 + \left(1 - \frac{\log v}{\log v_0} \right)^{-\delta^2/2} \partial^\mu \partial_\mu \right] \phi = 0 .
\end{equation}
The scalar $\phi$ should not be confused with the scalar that stabilizes the bulk geometry. We can transform the equation of motion into a Schr\"odinger form by rescaling $\phi$ as 
\begin{equation}
    \phi= (qv)^{3/2} \left(1 - \frac{\log v}{\log v_0} \right)^{-\delta^2/2} e^{iqt} \psi(v) .
\end{equation}
This yields the equation $(\partial_{v}^2-U(v))\psi(v)=0$, with the potential
\begin{equation}\label{eq:scalarpotential}
    U(v) = \frac{1}{v^2} \left[ \frac{15}{4} + \frac{2 \delta^2}{\log v_0/v} - \frac{\delta^2}{2} \left(1 - \frac{\delta^2}{2} \right) \frac{1}{\log^2 v_0/v}  \right] - \left(1 - \frac{\log v}{\log v_0} \right)^{-\delta^2/2} q^2 .
\end{equation}
This is similar to the procedure we followed for the shockwave equation of motion, c.f. Eqs.~\eqref{eq:shockwaveEOM} and~\eqref{eq:schrodingerEOM}.

Consider the high-energy regime $qv,qv_0\gg1$. Away from the singularity we can neglect the $1/v^2$ terms in the potential, so the equation of motion is just
\begin{equation}
    \left[\frac{1}{q^2}\partial_v^2 + \left(1 - \frac{\log v}{\log v_0} \right)^{-\delta^2/2} \right] \psi \sim \mathcal{O}\left(\frac{1}{q v} \right)  .
\end{equation}
Using Eq.~\eqref{eq:zderivatives} we see that this is solved by $\psi = e^{-iqz(v)}$ --- up to terms subleading in $1/qv$ --- yielding the high-energy wavefunction
\begin{equation}\label{eq:wavefunction}
    \phi \sim \left(\frac{v}{z(v)} \right)^{3/2}\left(1 - \frac{\log v}{\log v_0} \right)^{-\delta^2/2} (qz(v))^{3/2} e^{iq(t-z)} .
\end{equation}
This should be compared to the corresponding result in AdS, which is
\begin{equation}\label{eq:wavefunction_ads}
    \phi_{\rm AdS} \sim (q z)^{3/2} e^{iq(t-z)} .
\end{equation}
We remark that the extra terms appearing in $\phi$ but not $\phi_{\rm AdS}$ are a function of $v/v_0$.

All of these computations are in the original Poincar\'e coordinates, before performing the transformation mapping the celestial sphere to the null plane $x^+ = 0$. Upon performing that transformation and taking the limit $x^+ \rightarrow 0$, Eq.~\eqref{eq:wavefunction_ads} becomes a delta function $e^{i q x^-/2} \delta^2(x^\perp) \delta(z - 1)$.

Let us consider how this transformation affects the wavefunction Eq.~\eqref{eq:wavefunction}. Recall that when there is an explicit scale in the metric, it must also transform. In particular we take $z_0 \rightarrow z_0/x^+$, which preserves the ratio $z / z_0$ (or equivalently $v/v_0$). Hence the extra terms that appear in Eq.~\eqref{eq:wavefunction} but not the AdS wavefunction are unchanged by the transformation. Consequently the wavefunction is still localized at $x^\perp = 0, z = 1$ (in the high-energy limit). The overall normalization may not be the same as in AdS, but this is unimportant for our purposes.

\bibliographystyle{JHEP}
\bibliography{biblio}

\end{document}